\documentclass{aastex}
\usepackage{spr-astr-addons}
\usepackage{url}

\RequirePackage{color}

\begin{document}

\title{External Electromagnetic Fields of a Slowly Rotating
Magnetized Star with Gravitomagnetic Charge}
\shorttitle{Electromagnetic Field of NUT Star}
\shortauthors{Rahimov et al.}

\author{B.~J. Ahmedov\altaffilmark{1,2,3}}
\author{A.~V.~Khugaev\altaffilmark{1}}
 \and
\author{A.~A. Abdujabbarov\altaffilmark{1,2,4}} \email{ahmadjon@astrin.uz}

\altaffiltext{1}{Institute of Nuclear Physics,
        Ulughbek, Tashkent 100214, Uzbekistan}
\altaffiltext{2}{Ulugh Begh Astronomical Institute,
Astronomicheskaya 33, Tashkent 100052, Uzbekistan}

\altaffiltext{3}{Inter-University Centre for Astronomy \&
Astrophysics, Post Bag 4, Pune 411007, India}

\altaffiltext{4}{ZARM, University Bremen, Am Fallturm, 28359
Bremen, Germany}

\begin{abstract}

We study Maxwell equations in the external background spacetime of
a slowly rotating magnetized NUT star and find analytical
solutions for the exterior electric fields after separating the
equations of electric field  into angular and radial parts in the
lowest order approximation. The star is considered isolated and in
vacuum, with dipolar magnetic field aligned with the axis of
rotation. The contribution to the external electric field of star
from the NUT charge is considered in detail.

\end{abstract}

\section{Introduction}\label{intro}

The existence of strong electromagnetic fields is one of the most
important features of rotating magnetized neutron stars observed
as pulsars~\citep{hewish} and magnetars~\citep{dt92}. It was shown
starting pioneering paper of~\cite{d55} that the electric field is
induced due to the rotation of highly magnetized star. The general
relativistic effect of dragging of inertial frames is a source of
additional electric field around rotating magnetized relativistic
stars~\citep{mh97,mt92,ram01a,ram01b,ra04,japan}.

One of the exotic  solution of the Einstein's equation of general
relativity is achieved by introducing an extra nontrivial
parameter, the so-called gravitomagnetic monopole moment or NUT
charge.  The generalized solution describing spacetime of a
localized stationary and axisymmetric object with nonvanishing
gravitomagnetic charge is known as the Kerr-Taub-NUT, where NUT
stands for Newman-Unti-Tamburino~\citep{nut63}. In the presence of
the NUT charge the spacetime loses its asymptotically flatness
property and, in contrast to the Kerr spacetime, becomes
asymptotically nonflat. One of the features of the spacetime with
NUT charge is that the later  has no curvature singularity, there
are conical singularities on its axis of symmetry that result in
the gravitomagnetic analogue of Dirac's string quantization
condition~\citep[see e.g.,][]{misn63,misn68}. The conical
singularities can be removed by imposing an appropriate
periodicity condition on the time coordinate. However, this
generates  closed timelike curves in the spacetime that makes it
hard to interpret the solution as a regular black hole. In an
alternative interpretation, one may consider the conical
singularities as the source of a physical string threading the
spacetime along the axis of symmetry~\citep{b69}. In spite of
these undesired features, the NUT solution still serves as an
attractive example of spacetimes with asymptotic nonflat structure
for exploring various physical phenomena in general relativity.

At present there is no any observational evidence for the
existence of gravitomagnetic monopole or {\it magnetic mass}.
Therefore it is interesting to study the electromagnetic fields in
NUT space with the aim to get new tool for studying new important
general relativistic effects which are associated with nondiagonal
components of the metric tensor and have no Newtonian
analogues~\citep[see, e.g.][where solutions for electromagnetic
waves and interferometry in spacetime with NUT parameter have been
studied]{zonoz07,kkl08,ma08}. Kerr-Taub-NUT spacetime with Maxwell
and dilaton fields is also recently investigated
in~\citep{aliev08}. In our preceeding papers~\cite{mak08,aak08} we
have studied the magnetospheric structure and related effects in a
plasma surrounding a rotating, magnetized neutron star and charged
particle motion around compact objects in the presence of the NUT
parameter.  General relativistic effects associated with the
gravitomagnetic monopole moment of a gravitational source have
been investigated by~\citet{ak06} through the analysis of the
motion of test particles and electromagnetic fields distribution
in the spacetime around the nonrotating cylindrical NUT source.
The collision of two particles with the different rest masses
moving in the equatorial plane of a Kerr-Taub-NUT spacetime has
been considered by~\citet{liu11}. The analytic solutions of
Maxwell equations for infinitely long cylindrical conductors with
nonvanishing electric charge  and currents in the external
background spacetime of a line gravitomagnetic monopole have been
presented by~\citet{ak05}.

Here we study the exterior electric field produced in the
space-time of slowly rotating NUT star with dipolar magnetic field
aligned along axis of rotation.

This paper is organized as follows: in Section~\ref{potential} the
description of the model as well as derivation of the field
equation is presented. The magnetic moment of the star is
considered as non-varying in time as a result of infinite
conductivity. In Section~\ref{nutsolut} the analytical solution
for the electromagnetic field due to nonvanishing gravitomagnetic
charge is presented. We discuss obtained results and features of
the spacetime in Section~\ref{conclusion}. Throughout, we use a
space-like signature $(-,+,+, +)$ and a system of units in which
$G = 1 = c$, Greek indices run from 0 to 3, Latin indices from 1
to 3.

\section{Electromagnetic Field Equations Around Magnetized Star with NUT parameter}
\label{potential}

Our approach is based on the reasonable assumption that the metric
of spacetime is known, i.e. neglecting the influence of the
electromagnetic field on the gravitational one and finding
analytical solutions of Maxwell equations on a given, fixed
background. The next our approximation is in the specific form of
the background metric in a spherical coordinate system
$(ct,r,\theta,\phi)$ which we choose to be that of a stationary,
axially symmetric system truncated at the first order in the
angular velocity $\Omega$ and in gravitomagnetic monopole moment
$l$ as (see, for example,~\cite{dt02,betal03})
\begin{eqnarray}
\label{slow_rot} &&ds^2 = -N^2 dt^2 + N^{-2}dr^2 + r^2 d\theta ^2+
r^2\sin^2\theta d\phi ^2 \nonumber\\&&\qquad\quad - 2\left[\omega
(r) r^2\sin^2\theta + 2lN^2\cos\theta\right] dt d\phi \ ,
\end{eqnarray}
where parameter $N\ \equiv \left(1-{2 M}/{r}\right)^{1/2}$ and 
$\omega(r)\equiv{2J}/{r^3}$ can be interpreted as the angular
velocity of a
    free falling (inertial) frame and is also known as the
Lense-Thirring angular velocity, $J=I(M,R)\Omega$ is the total
angular momentum of metric source with total mass $M$. The
nondiagonal component of the metric tensor is finite at the
infinity: $\lim_{r\rightarrow\infty} g_{\rm 03}=-2l\cos\theta$
which is meant that the metric~(\ref{slow_rot}) is not
asymptotically flat.

Here we will look for stationary solutions of the Maxwell
equation, i.e. for solutions in which we assume that the magnetic
moment of the star does not vary in time as a result of the
infinite conductivity of the stellar interior (see for the details
of the stellar model to~\cite{af08}).
 We look for separable solutions of Maxwell equations in
the form
\begin{eqnarray}
B^{\hat r}(r,\theta) &=& F(r)\cos\theta\ , \nonumber\\ B^{\hat
\theta}(r,\theta) &= & G(r)\sin\theta\ , \nonumber\\  B^{\hat
\phi}(r,\theta) &=& 0\ ,
\end{eqnarray}
\noindent where functions $F(r)$ and $G(r)$ will account for the
relativistic corrections due to a curved background spacetime. In
the case of infinite conductivity and as far as the stationary
magnetic field is concerned, the study of Maxwell equations in a
slow rotation metric provides no additional information with
respect to a non-rotating metric. The dependence from the frame
dragging effects and gravitomagnetic charge is therefore expected
to appear at ${\mathcal O}(\omega^2,l^2)$. The stationary vacuum
magnetic field external to an aligned magnetized relativistic star
is known and given by~\cite{go64}
\begin{eqnarray}
\label{sol_mfe_1} && B^{\hat r} = - \frac{3\mu}{4M^3}
    \left[\ln N^2 + \frac{2M}{r}\left(1 +  \frac{M}{r}
    \right) \right] \cos\theta
    \ ,
\\\nonumber\\
\label{sol_mfe_2} && B^{\hat \theta} = \frac{3\mu N}{4 M^2 r}
    \left[\frac{r}{M}\ln N^2 +\frac{1}{N^2}+ 1
    \right] \sin\theta\ ,
\end{eqnarray}
where $\mu$ is magnetic dipole. The search for the form of the
electric field is much more involved than for the magnetic field.
However, hereafter we will make use of the insight gained
in~\cite{ram01a} as a guide and start the derivation of the
solution by rewriting vacuum Maxwell equations as
\begin{eqnarray}
\label{vac_1} && {\sin\theta}\left({\omega} r^2 B^{\hat
r}\right)_{,r}
    +\frac{2l\cos\theta}{\sin\theta}\left(N^2B^{\hat r}\right)_{,r}
    \nonumber\\
&& + {\omega} N^{-1}r\left(\sin\theta
    B^{\hat \theta}\right)_{,\theta}+ \frac{2lN}{r} \left(\frac{\cos\theta}{\sin\theta}
    B^{\hat\theta} \right)_{,\theta}  \nonumber\\
   && = \left(rN E^{\hat \theta}\right)_{,r}
    - E^{\hat r}_{\ ,\theta} \ ,
\\
&&\label{vac3}
     N\sin\theta\left(r^2 E^{\hat r} \right)_{,r}+
    {r}\left(\sin\theta E^{\hat \theta}\right)_{,\theta}
    + rE^{\hat \phi}_{\ ,\phi}
    = 0 \ ,
\end{eqnarray}

\noindent and which already indicate that the dragging of inertial
frames with angular velocity $\omega$ and gravitomagnetic monopole
$l$ introduce electric fields in the surrounding space when
magnetic fields are present. The exact solution for the vacuum
electric field can be represented by the sum
$E=E_{\Omega}+E_{\omega}+E_{l}$, where the subscripts $\Omega,
\omega, l$ indicate the fields generated by the rotation, the
dragging of inertial frames and gravitomagnetic charge,
respectively.
 Using as a reference
the solutions for slowly rotating magnetized sphere~\cite{ram01a},
we look for the simplest solutions of vacuum Maxwell equations in
the form
\begin{eqnarray}
\label{eeff_1}  &&\hspace{-0.7cm}E^{\hat r} = \Bigg\{
    \frac{15\omega r^3}{16M^5c}\bigg\{C_3\bigg[
    \left(3-\frac{2r}{M}\right)
    \ln N^2 + \frac{2M^2}{3r^2}-4
    \nonumber\\
    && \hspace{-0.7cm}+\frac{2M}{r}\bigg]+\frac{2M^2}{5r^2}\ln N^2
    +\frac{4M^3}{5r^3} \bigg\} + \frac{\Omega}{6 c R^2} C_1 C_2
    \bigg[\frac{2M^2}{3r^2}-4
    \nonumber\\&&\hspace{-0.7cm} +\left(3-\frac{2r}{M}\right) \ln N^2 + \frac{2M}{r}\bigg]
    \Bigg\}
    \left(3 \cos^2\theta-1\right)
    \mu +E^{\hat r}_{l} ,\ \ \
\\\nonumber \\
\label{eeff_2}  && \hspace{-0.7cm}E^{\hat \theta} = - \Bigg\{
    \frac{45\omega r^3}{16 M^5 c}N\bigg\{C_3
    \left[\left(1-\frac{r}{M}
    \right)\ln N^2-2-\frac{2M^2}{3r^2N^2}\right]
\nonumber\\
    &&\hspace{-0.7cm} +\frac{4M^4}{15r^4 N^2}\bigg\}
    + \frac{\Omega}{2 c R^2} C_1 C_2 N
    \bigg[\left(1-\frac{r}{M}
    \right)\ln N^2\nonumber\\&&\hspace{-0.7cm} -2-\frac{2M^2}{3r^2N^2}\bigg]
    \Bigg\}
    \mu\sin2\theta
    +E^{\hat \theta}_{l} .
\end{eqnarray}

    The values of the arbitrary constants $C_1$, $C_2$ and
$C_3$ are found after imposing the continuity of the tangential
electric field across the star surface~\cite{ram01a}
\begin{eqnarray}
\label{c1}  C_1 &=& - \frac{3R^3}{4M^3}
    \left[\ln\left(1-\frac{2M}{R}\right)
    +\frac{2M}{R}\left(1+\frac{M}{R}\right)\right]\nonumber\\
  &  = &\frac{F(R) R^3}{\mu}\ ,
\\ \nonumber \\
\label{c2}  C_2& =&
\frac{1}{N^2_{_R}}\left[\left(1-\frac{R}{M}\right)
    \ln N^2_{_R} - 2 - \frac{2 M^2}{3R^2 N^2_{_R}}\right]^{-1} \ ,
\\
   C_3 &= &\frac{2M^2}{15 R^2}C_2
    \left[\ln N^2_{_R} + \frac{2M}{R}\right]\ ,
\end{eqnarray}
with $N^2_{_R} \equiv N^2(r=R) = 1 - 2M/R$.

For unknown components $E_l^{\hat{r}}$ and $E_l^{\hat{\theta}}$ of
the electric field we have the following set of differential
equations
\begin{eqnarray}
\label{vacGM1}
  &&\sin\theta\left(r^2 E^{\hat r}_{l} \right)_{,r}+
    N^{-1}\left(\sin\theta E^{\hat \theta}_{l}\right)_{,\theta}
    = 0 \ ,\\ \nonumber\\
 &&\frac{3\mu
l}{M^2r^2}\bigg[\frac{\cos^2\theta}{\sin\theta}\left(\ln
N^2+\frac{2M}{r}+ \frac{2M^2}{r^2}\right)+
\frac{N^2\sin\theta}{2}\nonumber\\
 &&\times \left(\frac{r}{M}\ln N^2+\frac{1}{N^2}+1\right)\bigg]
    \nonumber\\&&-E^{\hat r}_{l\ ,\theta} +\left(rN E^{\hat
\theta}_{l}\right)_{,r} =0 \ , \label{vacGM2}
\end{eqnarray}
derived from the Maxwell equations (\ref{vac_1}) and (\ref{vac3}).

\section{Electromagnetic Field due to nonvanishing Gravitomagnetic Charge\label{nutsolut}}

In this section we will try to find exact analytical solution for
electromagnetic field of magnetized neutron star due to
nonvanishing gravitomagnetic charge. Introducing new variables
\begin{eqnarray}
\label{F}  F_1(r,\theta)&=&r^2E^{\hat{r}}_{l}\ , \nonumber\\
F_2(r,\theta)&=&rN\sin\theta E^{\hat{\theta}}_{l}\ ,\quad
\nonumber \\
 G(r,\theta)&=&\frac{3\mu
l}{M^2r^2}\left[\frac{\cos^2\theta}{\sin\theta}\left(\ln
N^2+\frac{2M}{r}+ \frac{2M^2}{r^2}\right) \right.\nonumber\\&&
\left. + \frac{N^2\sin\theta}{2}\left(\frac{r}{M}\ln
N^2+\frac{1}{N^2}+1\right)\right]\ ,
\end{eqnarray}
and making some algebraic transformations one can write the
equations~(\ref{vacGM1}) and (\ref{vacGM2}) in more convenient way
\begin{eqnarray}
\label{vacGMn1} &&
r^2N^2F_{1,rr}+F_{1,\theta\theta}+2MF_{1,r}+\nonumber\\
&& \cot\theta F_{1,\theta}-\frac{r^2}{\sin\theta}(\sin\theta
G)_{,\theta}=0\ ,
\\ \nonumber\\
\label{vacGMn2} && F_{2,\theta}=- N^2F_{1,r}\sin\theta\ .
\end{eqnarray}
The last term in the right hand side of the
equation~(\ref{vacGMn1}) in the linear in $l$ approximation and
neglecting $lM/r^2 \ll 1$ takes the following form
\begin{equation}
\frac{r^2}{\sin\theta}(\sin\theta G)_{,\theta}\approx \frac{4\mu
l}{r^2}\cos\theta +{\mathcal{O}}\left(\frac{lM}{r^2}\right)\ .
\end{equation}

Inserting these expansions into equation~(\ref{vacGMn1}) one can
get the set of equations to be solved including~(\ref{vacGMn2})
and the following one:
\begin{equation}
\label{vacGM_fin}
\xi\left(\xi-1\right)F_{1,\xi\xi}+F_{1,\theta\theta}+F_{1,\xi}+\cot\theta
F_{1,\theta}-\frac{\mu l}{M^2\xi^2}\cos\theta =0\ ,
\end{equation}
which has been rewritten in new variable $\xi ={r}/{2M}$.

Solution of equation~(\ref{vacGM_fin}) can be easily  presented in
the more convenient form, which has been rewritten in new
nondimensional variables
\begin{equation}
\label{vacGM_separat} \xi(\xi-1)f_{,\xi\xi}+f_{,\xi}- 2f=\frac{\mu
l}{M^2\xi^2}\ ,
\end{equation}
through two separate functions of different variables $\xi$  and
$\theta$ as $F_{1}(\xi,\theta)=f(\xi)\cos\theta$. The equation
(\ref{vacGM_separat}) can be solved analytically as
\begin{eqnarray}
F_1(\xi,\theta)&=&\frac{\mu l \xi^2 }{M^2}
\bigg\{C_4\Bigl[\frac{1}{\xi}+
\frac{1}{2\xi^2}\nonumber\\&&+\ln\Bigl(\frac{\xi-1}{\xi}\Bigr)\Bigr]
- \frac{1}{3\xi^3}\bigg\} \cos\theta\ ,
\end{eqnarray}
where $C_4$ is the integration constant.

According to the definition (\ref{F}) the values of electric field
$E_l^{\hat{r}}$ and $E_l^{\hat{\theta}}$ produced by the effect of
gravitomagnetic charge $l$  are
\begin{eqnarray}
\label{EF1}  E_l^{\hat{r}}(r,\theta)&=&\frac{\mu
l}{Mr^3}\Bigg[C_4\bigg(1+\frac{r} {M} \nonumber\\
&& +\frac{r^2}{2M^2}\ln N^2\bigg)\frac{r}{2M} -\frac{2}{3}\Bigg]
\cos\theta\ ,
\\ \nonumber\\
\label{EF2}  E_l^{\hat{\theta}}(\xi,\theta)&=& \frac{\mu l}{Mr^3}
\bigg[\frac{1}{6}-\left(1+ \frac{r}{M}+\frac{r^2}{M^2}N^2\ln
N\right)\nonumber\\&& \times
C_4\frac{r}{4M}-\frac{M}{3r}\bigg]N\cos2\theta .\ \ \ \quad
\end{eqnarray}
where $C_4$ can be easily found from boundary condition and has
the following form:
\begin{eqnarray}
&& C_4=\bigg[\frac{2M}{3R}\left(1-\frac{6M}{R}\right)-\frac{3R}{M
N_{R}} \left(1+N_{R}^{2}\right)\nonumber\\&& -
\frac{3R^2}{M^2}N_R\ln N_{R}^2\bigg]\left[1+\frac{R}{M}+\frac{R^2
}{M^2}N_{R}^{2}\ln N_R\right]^{-1} .\  \
\end{eqnarray}

\begin{figure*}
a) \includegraphics[width=0.45\textwidth]{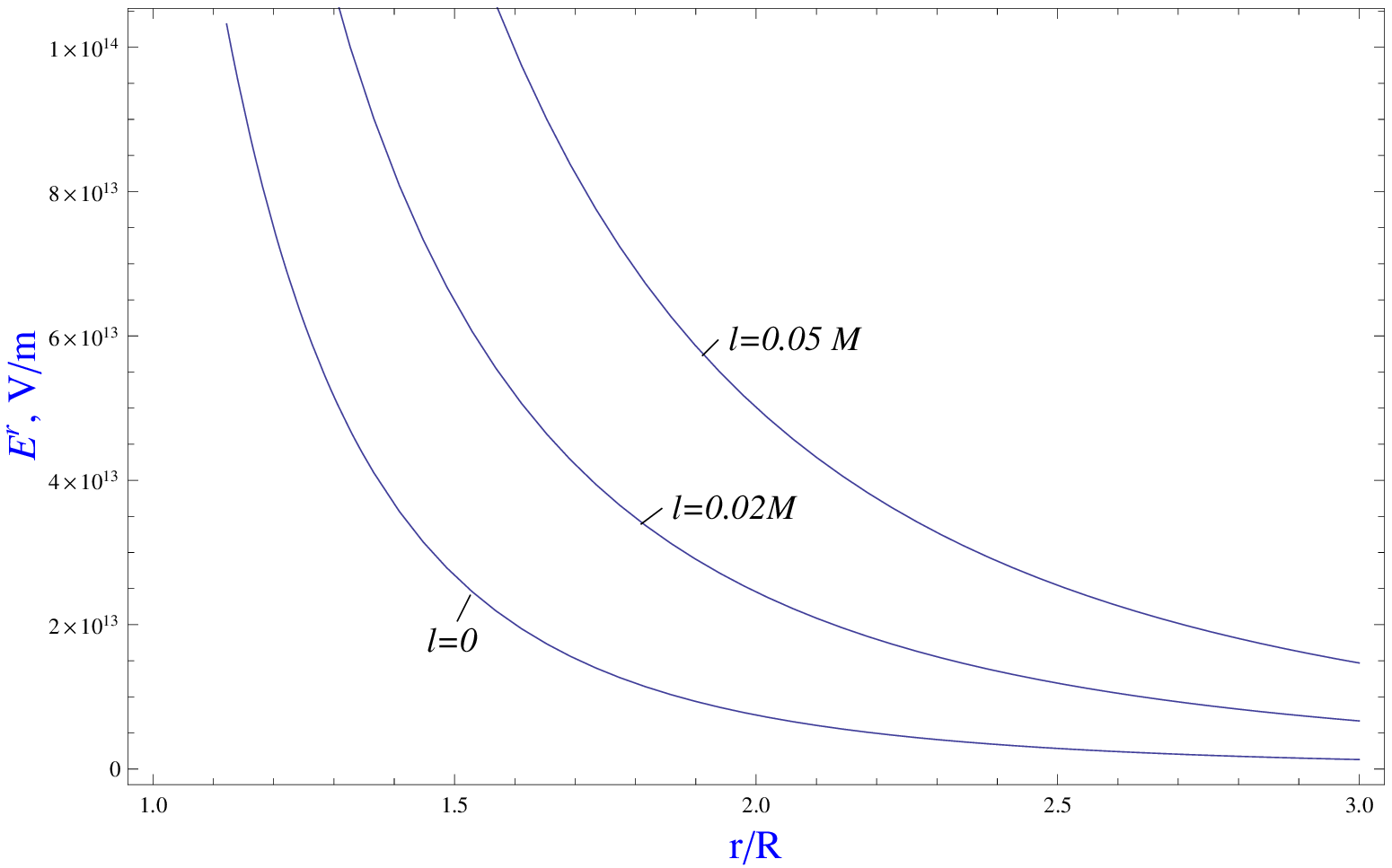}
b) \includegraphics[width=0.45\textwidth]{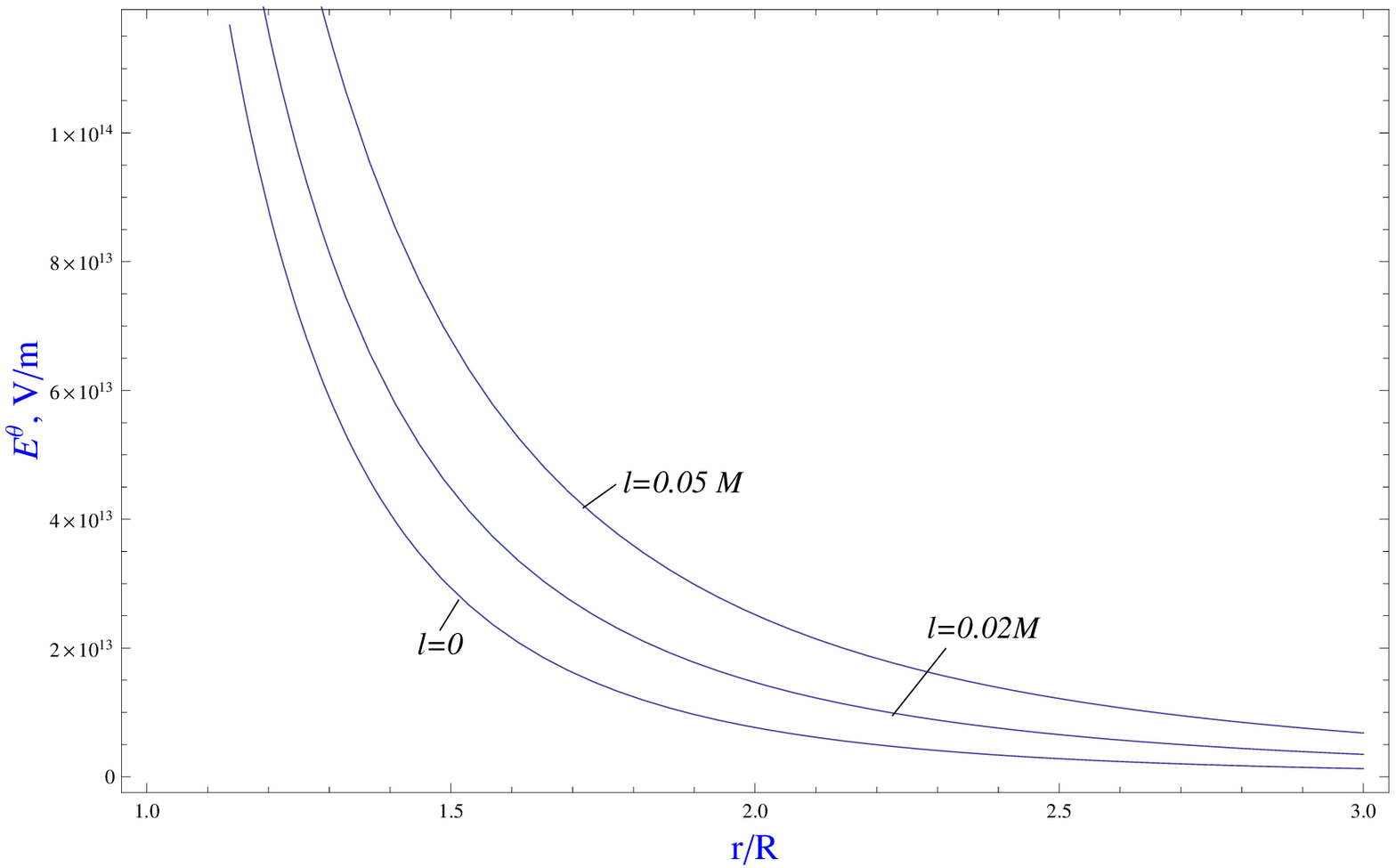}
\caption{\label{fig:1} Radial dependence of the radial (a) and
azimuthal (b) components of the electric field of magnetized
rotating star with nonvanishing NUT parameter for the different
values of gravitomagnetic monopole momentum $l$. {The units of the
gravitomagnetic monopole momentum are taken in mass of the star
with mass $M=1.5 M_\odot$}.}
\end{figure*}

For the models of neutron star in the present study we choose the
following parameters $R=10~km$ and the polar surface field
strength $B(\overline{r}=r/R=1)=10^{12}G$, period $T=10^{-2}$ s.
Following this, we plot the radial dependence of the external
electric field for various values of $l$ as shown in
Fig.~\ref{fig:1}. The enhancement of the exterior electric field
at the surface of the relativistic star is given in
Table~\ref{table1} which varies for several times depending on
parameter $l$ selected. As can be seen from the presented graphs
in Fig.~\ref{fig:1} an additional electric field produced by NUT
parameter is becoming extremely important outside the star
depending on the value of parameter $l$ selected and maybe more
relevant to observational phenomenology from slowly rotating
pulsars with low rotation period.

\begin{table*}
\caption{ \label{table1}The amplification of the magnitude of
exterior electric field around the typical relativistic compact
star with gravitomagnetic monopole momentum in the distance of $R$
from the surface of the star. The mass of star is
$M=1.5M_{\odot}$, period $T=10^{-2}$ s and the radius is
$R=10~km$. The values of NUT parameter  $l$ are taken as usual in
units of $M$, the values of electric field are taken in units
$~10^{13}~V/m$. {For the comparison we also provide the values for
the exterior electric field corresponding to the upper values of
the NUT charge obtained by \citet{habibi04,ma08}.}}
\begin{center}
{\begin{tabular}{@{}ccccccccc@{}} \hline\noalign{\smallskip}
 ${\bf l}$ & 0.006 & 0.01 &  0.02 & 0.03 & 0.05 & 0.08 & 0.1 & 4.52 \\
 & \citet{habibi04} &  &   &   & &   &   &  \citet{ma08} \\
\noalign{\smallskip}\hline\noalign{\smallskip}
 ${\bf E}$ & 0.977 & 1.14 & 1.61 & 2.12 & 3.18 & 4.81 & 5.90 & 249.74 \\
 \noalign{\smallskip}\hline
\end{tabular}}
\end{center}
\end{table*}
\section{\label{conclusion} Conclusion}

Here we have considered the external solution for electromagnetic
fields of magnetized neutron star in the presence of nonvanishing
NUT parameter. The central object considered as slowly rotating
and magnetic moment of the latter is considered not changing by
time. The dependence of the electromagnetic field from the NUT
charge in analytical way is presented in this paper. In the
previous works the upper limit for the gravitomagnetic charge has
been obtained comparing astrophysical data with theoretical
results as ({the units of the gravitomagnetic charge are taken in
units of stellar mass $M=1.5 M_\odot$})(i) $l\leq 0.006 M$ from
the gravitational microlensing \citep{habibi04}, (ii) $l\leq 4.52
M$ from the interferometry experiments on ultra-cold atoms
\citep{ma08}, (iii) and similar limit has been obtained from the
experiments on Mach-Zehnder interferometer \citep{kkl08}. In the
linear approximation in NUT parameter the external magnetic field
depends only from mass of neutron star whereas exterior electric
fields depend on the gravitomagnetic charge linearly. The detailed
comparison of the astrophysical data from pulsars related the
electromagnetic phenomena might give one more precise
astrophysical limit for the NUT parameter in this framework.
\section*{\label{acknow} Acknowledgments}

Authors thank the IUCAA for warm hospitality during their stay in
Pune and AS-ICTP for the travel support. This research is
supported in part by the UzFFR (projects 1-10 and 11-10) and
projects FA-F2-F079 and FA-F2-F061 of the UzAS. ABJ thanks the
TWAS for the Regular Associateship grant. AAA thanks the German
Academic Exchange Service (DAAD) for financial support.

\end{document}